\documentclass[fleqn,usenatbib]{mnras}

\usepackage{lipsum}
\usepackage{mwe}

\usepackage{lipsum,graphicx}
\usepackage{newtxtext,newtxmath}
\usepackage[T1]{fontenc}
\usepackage{ae,aecompl}
\usepackage{adjustbox}
\usepackage{amsfonts}
\usepackage{amsmath}
\usepackage{graphicx}	
\usepackage{tabularx}
\usepackage{verbatim}
\usepackage{xspace}
\usepackage{adjustbox}
\usepackage{rotating, lipsum, babel}
\usepackage{MnSymbol}

\usepackage{subfig}

\newcommand{\rh}{r_\mathrm{h}}
\newcommand{\RG}{R_\mathrm{G}}

\newcommand{\rt}{r_\mathrm{t}}

\newcommand{\mm}{\mathrm{<\!m\!>}}

\newcommand{\floss}{f_\mathrm{mass-loss}}

\NewDocumentCommand{\Msun}{o}{
  \IfNoValueTF{#1}
    {\,\mathrm{M}_{\odot}}
    {10^{#1}\,\mathrm{M}_{\odot}}
}

\newcommand{\pc}{\,\mathrm{pc}}
\newcommand{\kpc}{\,\mathrm{kpc}}

\newcommand{\secref}[1]{Section~\ref{#1}}
\newcommand{\figref}[1]{Figure~\ref{#1}}
\newcommand{\tabref}[1]{Table~\ref{#1}}

\newcommand{\ltail}{l_\mathrm{tail}}
\newcommand{\rcm}{r_\mathrm{cm}}

\title[The fingerprint of PMS on tidal tails]{The fingerprint of primordial mass segregation on the tidal tails of star clusters}

\author[Ghasemi et al.]
{S. Mojtaba Ghasemi$^{1}$, Hosein Haghi$^{1,2,3,4}$\thanks{
E-mail:  \mbox{haghi@iasbs.ac.ir} (HH)},
Mohammad Mansoury$^{1}$,
Akram Hasani Zonoozi$^{1,2}$,
\newauthor Pouriya Miri$^{1,5}$,
and Pavel Kroupa$^{2, 6}$\\
$^{1}$Department of Physics, Institute for Advanced Studies in Basic Sciences (IASBS), PO Box 11365-9161, Zanjan, Iran\\
$^{2}$Helmholtz-Institut f\"ur Strahlen-und Kernphysik (HISKP), Universit\"at Bonn, Nussallee 14-16, D-53115 Bonn, Germany\\
$^{3}$School of Astronomy, Institute for Research in Fundamental Sciences (IPM), PO Box 19395 − 5531, Tehran, Iran\\
$^{4}$Center for International Scientific studies and Collaborations (CISSC), Ministery of Science, Research and Technology, PO Box 15986-86515, Tehran, Iran\\
$^{5}$Faculty of Computer and Information Science,  University of Ljubljana, večna pot 113, 1000 Ljubljana, Slovenia\\
$^{6}$Charles University in Prague, Faculty of Mathematics and Physics, Astronomical Institute, V Hole\v{s}ovi\v{c}k\'ach 2, CZ-180 00, 
Praha 8, Czech Republic\\
}

\date{Accepted XXX. Received YYY; in original form ZZZ}

\pubyear{2025}

\begin{document}
\label{firstpage}
\pagerange{\pageref{firstpage}--\pageref{lastpage}}
\maketitle

%%%%%%%%%%%%%%%%%%%%%%%%%%%%%%%%%%%%%%%%%%%%%%%%%%%%%%%%%%%%%%%%%%%%%
%%%%%%%%%%%%%%%%%%%%%%%%%%%%%%%%%%%%%%%%%%%%%%%%%%%%%%%%%%%%%%%%%%%%%

\begin{abstract}
 
We investigate the effect of primordial mass segregation (PMS) in shaping the tidal tail structures of star clusters, searching for any trace of PMS on the tails at both early and late evolutionary stages. Through N-body simulations, we analyze clusters with two different degrees of PMS at various Galactocentric distances ($\RG$), considering two black hole retention scenarios. Our findings reveal that PMS influences early cluster expansion and the formation of tidal tails with a bottom-heavy stellar mass function, this being more pronounced at smaller $\RG$ but diminishes over time. Primordially segregated clusters exhibit denser, unified, and longer tail structures compared to non-segregated clusters. The mean stellar mass distribution along the tails shows distinct patterns for primordially segregated and non-segregated clusters, converging at later evolutionary stages. The retention of stellar remnants has a weak impact on the mean mass distribution along the tails and on its morphology. We find that although mean mass differences persist along the tidal tails, the rate of change in primordially mass-segregated clusters eventually converges with that of non-segregated clusters, suggesting that the influence of primordial mass segregation on the tidal tails gradually diminishes over the course of cluster evolution.

\end{abstract}


%%%%%%%%%%%%%%%%%%%%%%%%%%%%%%%%%%%%%%%%%%%%%%%%%%%%%%%%%%%%%%%%%%%%%
\section{Introduction}\label{sec:intro}

Mass segregation, the process by which heavier stars sink toward the centre while lighter stars move outward on a timescale proportional to the two-body relaxation time, is a natural consequence of the energy equipartition in stellar systems such as star clusters \citep{spitzer2014dynamical, bonnell1998}. This so-called dynamical mass segregation may be accompanied by primordial mass segregation (PMS), which occurs during the star formation phase and significantly influences the subsequent dynamical evolution of star clusters, including the distribution of stars in their growing tidal tails. Though the exact mechanisms driving PMS remain uncertain, they are believed to be linked to the star formation process itself \citep{murray1996, bonnell2002, mcmillan2007}. The degree of PMS can be quantified using the segregation coefficient \( S \), which measures the correlation between stellar mass and radial position and ranges from $S=0$ (no segregation) to $S=1$ (fully segregated). Observations of some young clusters (e.g., the Orion Nebula Cluster, NGC 330, etc) revealed $S$ values typically between 0.2 and 0.9, with some clusters reaching $S \geq 0.8$, suggesting significant PMS in their early stages \citep{hillenbrand1997, fischer1998, sirianni2002, gouliermis2004, stolte2006, sabbi2008, allison2009, gouliermis2009, de2010, Pavlik2019}.  

Regardless of its origin, PMS plays a crucial role in shaping the long-term evolution of star clusters. In tidally filled clusters, PMS can lead to their enhanced expansion due to early massive star evolution, increasing mass loss across the tidal boundary, and accelerating cluster dissolution \citep{haghi2014}. In contrast, tidally under-filling clusters can better retain their structure despite this early expansion \citep{Vesperini2009PMS}. PMS also affects the early gas-expulsion phase and influences the evolution of the stellar mass function (MF). In particular, PMS may produce a bottom-heavy MF in the outer regions of clusters at early evolutionary stages. As massive stars sink toward the cluster core while low-mass stars are preferentially lost through evaporation and tidal stripping, dynamically evolved clusters with PMS can develop flatter core MFs compared to initially non-segregated clusters (e.g., \citealt{Zonoozi2011, Zonoozi2014,  Zonoozi2017, Zonoozi2024, Haghi2015}). This raises an important question: Are embedded star clusters born mass-segregated, and if so, how strongly?  

Studies of outer-halo globular clusters (GCs) like Pal 4 and Pal 14 provide intriguing evidence for PMS, showing flat stellar mass functions with a deficiency of low-mass stars and unexpected mass segregation at their centres, despite two-body relaxation times exceeding a Hubble time \citep{Jordi2009, Frank2012, Frank2014}. This suggests that these clusters may have been born compact, with PMS shaping their early evolution \citep{Zonoozi2011, Zonoozi2014, Zonoozi2017}.  

As star clusters evolve dynamically within their host galaxy, they lose stars primarily through external tidal effects, such as tidal stripping and tidal shocks \citep{Gnedin1997, Weatherford2023}, while two-body relaxation-driven evaporation also contributes to the gradual escape of stars. These processes can lead to the formation of tidal tails, i.e., long and thin stellar streams extending outward from the cluster. These structures result from energy-equipartition-driven stellar evaporation and provide a powerful tool for studying both cluster evolution and the Galactic gravitational potential. Due to their low-velocity dispersion, GC tidal streams are particularly sensitive to small gravitational perturbations, making them valuable tracers of the Milky Way's potential (e.g., \citealt{Kupper2008, Kupper2010, Kupper2012, Grillmair2013, Bernard2014, Koposov2015}). The abundance of observed thin streams offers crucial insights into the shape of the Milky Way’s gravitational field \citep{Bonaca2014}.  

%%%%%%%%%(e.g., Bonaca, Geha \& Kallivayalil 2012, Grillmair et al. 2013, Bernard et al. 2014, Koposov et al. 2014)   (Bonaca et al. 2014)  ^^^^^^^^

The study of tidal tails dates back to early theoretical predictions by \citet{Bok1934} and \citet{Spitzer1940} and has since been advanced through extensive observational \citep{Odenkirchen2001, Grillmair2009, Keller2009, Ibata2019, Kos2024} and numerical \citep{Baumgardt2003, Montuori2007, Chumak2010, Kupper2010, Kupper2012, Jerabkova2021, Boffin2022, wang2021, Jadhav2025} investigations.  Observationally, Palomar 5 was among the first GCs to exhibit pronounced tidal tails, extending more than 10 degrees in the sky \citep{Grillmair1995, Odenkirchen2003}. Similar structures have been identified in NGC 288 \citep{Shipp2018, Kaderali2019NGC288}, NGC 5466, and NGC 5053 \citep{Lauchner2006, Belokurov2006}. By introducing the innovative compact convergent point method \citet{Jerabkova2021} were able to map the complete extend of the tidal tails of the Hyades open cluster.   The Hyades cluster exhibits extended tidal tails up to 800 pc in length \citep{Meingast2019, Roser2019}, revealing the complex spatial distribution of escaped stars.  

Theoretical and numerical studies have deepened our understanding of the formation and evolution of tidal tails. \citet{Montuori2007} demonstrated through N-body simulations that tidal tails arise from interactions with the gravitational field of the host galaxy. Delving deeper into the area of tidal tail simulations, \citet{Webb2022} studied the behaviour of the MF slope along the tidal tails of star clusters. Their analysis revealed that stellar streams originating from star clusters exhibit variations in their stellar MF along the streams of star clusters that dissolved after several relaxation times. At the edges of the stream, the MF remains near primordial due to early escapees. Moving inward, the MF steepens as lower-mass stars that escaped during mass segregation dominate. Near the stream's centre, the MF flattens again due to the high-mass stars escaping last. In addition, attempting to replicate the observational data, \citet{Gieles2021Pal5Sim} explored the dynamical history and future evolution of the tidal tails of Palomar 5. Additionally, kinematically cold GC streams provide valuable constraints on the Galactic potential, as their stars exhibit epicyclic motion upon escape from the cluster leading to a series of regularly spaced tail overdensities \citep{Kupper2008, Just2009, Kupper2010, Kupper2012}, noted also by \citet{Capuzzo2005, Montuori2007}.

Recent Gaia data have revealed another important feature: Tidal tails can be strongly asymmetric between the leading and trailing arms. \citet{Kroupa2022} and \citet{Kroupa2024} found that several nearby open clusters show a significant excess of stars in their leading tails, a result difficult to reproduce with Newtonian gravitation but consistent with Milgromian dynamics \citep{Pflamm-Altenburg2023}.  

At the same time, unresolved binaries have been shown to bias inferred cluster properties. \citet{Wirth2024} demonstrated that the presence of unresolved binaries leads to systematic underestimates of cluster mass and produces artificially shallow stellar mass functions in both clusters and their tidal tails. This highlights the importance of accounting for binary fractions when interpreting MF slopes in streams.

Beyond their role in Galactic dynamics, thin GC streams offer insights into the initial conditions of GCs at birth. Given the observational evidence for PMS in young star clusters and dynamically evolved GCs, it is likely that at least some, if not all, GCs formed with PMS. Since dynamically evolved clusters tend to have shallower MF slopes, their tidal streams are expected to contain a higher proportion of low-mass stars. This raises the question of how PMS affects the stellar MF of tidal tails over time.  

In this study, we investigate the impact of PMS on the evolution of tidal tails using simulation models with varying degrees of PMS. Specifically, we analyze the mean stellar mass (\(\mm\)) and MF slope (\(\alpha\)) across different spatial regions of the tidal tails. By comparing our simulation results with observational data from clusters such as those studied by \citet{Malhan2018A, Malhan2018B, Palau2019}, we aim to establish a connection between the primordial conditions of star clusters and the present-day structure of their tidal tails. Our ultimate goal is to infer primordial mass segregation coefficients for observed clusters based on their tidal tail properties.  The paper is organized as follows: In \secref{sec:method}, we describe our simulation methodology and initial conditions. \secref{sec:results} presents the results on the evolution of cluster parameters and their implications.

%%%%%%%%%%%%%%%%%%%%%%%%%%%%%%%%%%%%%%%%%%%%%%%%%%%%%%%%%%%%%%%%%%%%%
%%%%%%%%%%%%%%%%%%%%%%%%%%%%%%%%%%%%%%%%%%%%%%%%%%%%%%%%%%%%%%%%%%%%%

\section{Models and initial conditions}\label{sec:method}

The simulations of model star clusters were carried out using the NBODY7 code \citep{aarseth1999, aarseth2003}, which employs advanced algorithms for direct N-body integration. This code incorporates an algorithmic regularization chain, enhancing its ability to handle multiple systems in dense environments \citep{kS1965, Mikkola1993, Mikkola1999}. Additionally, the integration of single and binary star evolution via the SSE and BSE libraries \citep{hurley2000, Hurley2002} allows for  modelling of stellar evolution, including mass loss and remnant formation.  

For all models, the initial density profile and velocity distribution follow the \citet{plummer1911} profile (also see \citealt{Heggie2003, Kroupa2008}). Each simulated cluster starts with an initial mass of \(3\times10^4\Msun\) and a half-mass radius of \(\rh = 3\) pc, with no primordial binaries. The initial stellar masses are drawn from the \citet{kroupa2001variation} initial mass function (IMF), assuming a stellar metallicity of 0.005. The IMF thus follows the canonical two-part power-law distribution of stellar masses:

\begin{equation}
\frac{\mathrm{d}N(m)}{\mathrm{d}m} \propto
\begin{cases}
m^{-1.3} & 0.07 \leq \frac{m}{\Msun} < 0.5   \\
m^{-2.3} & 0.5 \leq \frac{m}{\Msun} < 150,
\end{cases}
\label{IMF}
\end{equation}
where, $\frac{\mathrm{d}N(m)}{\mathrm{d}m}$ is the number of stars with masses between $m$ and $m+dm$. Given the initial mass and IMF, the total initial number of stars is approximately $55000$ for our models.

\begin{figure}
   \centering
    \includegraphics[width=0.99\linewidth]{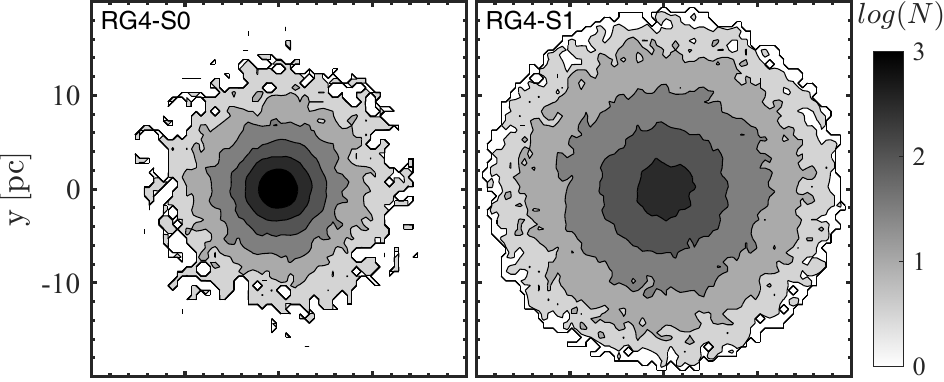}
   \includegraphics[width=0.99\linewidth]{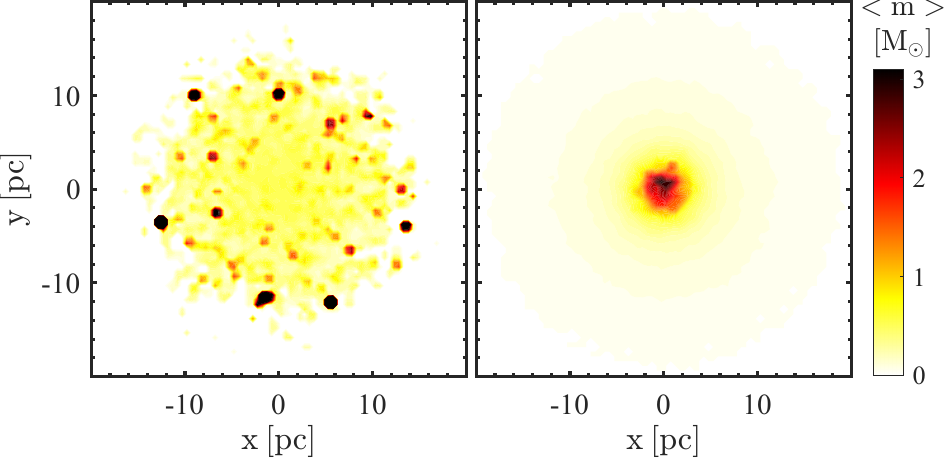}
    \caption{2-D display of the initial spatial distribution of stellar population characteristics. Top row: The logarithmic number density of stars. Bottom row: The distribution of mean stellar mass. Models with low and high PMS coefficients are shown in the left and right columns, respectively. Colour scales represent values per 2$\pc^2$ area.}
    \label{fig:initial-MeanMass-contour}
\end{figure}

The degree of primordial mass segregation is characterized by the parameter $S$ \citep{Suber2008}, which describes the correlation between stellar mass and radial position, such that $S=0$ corresponds to a non-segregated cluster, and $S=1$ refers to full segregation, where the most massive star is in the lowest point of the cluster potential and the least massive star is in the highest point of the cluster potential. To explore the impact of primordial mass segregation (PMS), we consider two distinct initial star distributions: one with no initial mass segregation (\(S = 0\), hereafter S0) and another with extreme segregation (\(S = 0.95\), hereafter S1). The initial distribution of mean stellar mass and number density for these two cluster models is illustrated in \figref{fig:initial-MeanMass-contour}.  

The star clusters are embedded in a static galactic potential consisting of three components \citep{aarseth2003, kupper2011}: a central bulge, a disk, and a phantom dark matter halo, ensuring a circular velocity of $220 \rm ~km/s$ at \( R_{\rm G} = 8.5 \) kpc. The initial orbital velocities of the clusters are assigned to maintain circular orbits at their respective Galactocentric distances (\(\RG\)).  

To investigate the effect of tidal fields, each cluster model is placed in three distinct Galactic orbits at 4, 12, and 30 kpc from the Galactic centre, covering a broad range of tidal-filling factors. Massive stars receive minimal natal kick velocities (NKVs) of approximately 1 km/s upon evolving into supernova remnants. Additionally, to assess the influence of NKVs, we repeat the simulations at \( \RG = 4 \) kpc with high NKVs, assuming a Gaussian natal kick velocity distribution for black holes and neutron stars with a dispersion of (\(\sigma_\mathrm{BH}\)) of 190 km/s. These models are denoted with the "-K" suffix.  Our goal is to examine how early mass segregation influences the dynamical evolution of star clusters and their tidal tails. For clarity, the initial parameters used in the simulations are summarized in \tabref{table:all}.

\begin{table*}
    \caption{The initial parameters that define our simulated clusters, including Galactocentric distances and PMS coefficients. The $t_{(\floss)}$ shows the time in which each model reaches its indicated $\floss$ milestone. All models begin with identical initial masses ($30 \times 10^{3} \Msun$) and half-mass radii ($\rh = 3\pc$).}
	\centering
	\begin{tabular}{ccccccccc}
		
  		\hline 		
            Model & $R_G$ & $S$ & $\sigma_\mathrm{BH}$ & & & $t_{(\floss)} \mathrm{[Gyr]}$ & & \\
            & $\mathrm{[kpc]}$ & & $\mathrm{[km \; s^{-1}]}$ & & (0.25) & (0.4) & (0.75) & (0.9) \\
            \hline
            
            RG4-S0     & 4   & 0.00 & 1   & & 0.14   & 0.70   & 2.75     & 3.28     \\
            RG4-S1     & 4   & 0.95 & 1   & & 0.14   & 0.57   & 1.73     & 2.05     \\
            RG4-S0-K   & 4   & 0.00 & 190 & & 0.05   & 0.38   & 3.07     & 4.82     \\
            RG4-S1-K   & 4   & 0.95 & 190 & & 0.05   & 0.29   & 1.76     & 2.63     \\
            RG12-S0    & 12  & 0.00 & 1   & & 0.22   & 1.80   & 7.50     & $>$ 13.6 \\
            RG12-S1    & 12  & 0.95 & 1   & & 0.25   & 1.52   & 6.82     & 8.46     \\
            RG30-S0    & 30  & 0.00 & 1   & & 0.24   & 3.10   & $>$ 13.6 & $>$ 13.6 \\ 
            RG30-S1    & 30  & 0.95 & 1   & & 0.28   & 2.94   & $>$ 13.6 & $>$ 13.6 \\

            \hline
        
	\end{tabular}
	\label{table:all}
\end{table*}

%+++++++++++++++++++++++++++
%%%%%%%%%%%%%%%%%%%%%%%%%%%%%%%%%%%%%%%%%%%%%%%%%%%%%%%%%%%%%%%%%%%%%
%%%%%%%%%%%%%%%%%%%%%%%%%%%%%%%%%%%%%%%%%%%%%%%%%%%%%%%%%%%%%%%%%%%%%

\section{Results}\label{sec:results}

Primordially mass-segregated star clusters expand rapidly in their early evolution due to stellar evolution processes in their dense cores \citep{Vesperini2009PMS,haghi2014}. This expansion is influenced by the retention of stellar remnants (especially BHs) where higher BH retention leads to a more rapid increase in cluster size, ultimately accelerating cluster dissolution under tidal forces \citep{Ghasemi2024}. Clusters with high PMS also have more low-mass stars in their outer regions, whose high velocity dispersions contribute significantly to the formation of extended tidal tails.

The Galactocentric distance ($\RG$) significantly affects cluster evolution. Clusters at larger $\RG$ experience greater expansion and slower mass loss, leading to more stable structures and broader tidal tails.
To analyse the impact of $\RG$ and PMS on cluster dynamics and tidal tail structures, we conducted comprehensive comparative analyses across various star cluster models. Clusters were examined at equivalent mass loss fractions for consistency. Our study encompasses simulations of eight distinct star cluster models (\tabref{table:all}), each characterised by two different degrees of mass segregation, $S$, and positioned at three different $\RG$. Also, to measure the effect of the BH retention, two star cluster models at $\RG = 4 \kpc$ are replicated with high NKV, which is equal to almost no initial retention of BHs. The following sections present a detailed description of the results derived from these simulations.

\begin{figure*}
    \includegraphics[width=0.87\linewidth]{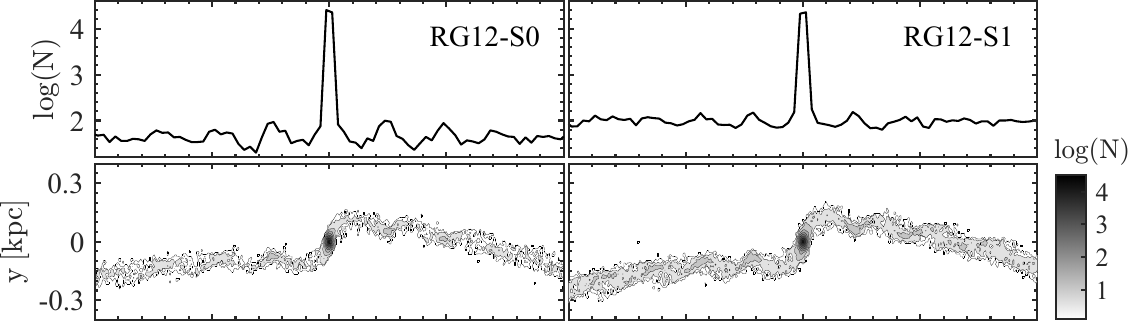}
    \includegraphics[width=0.87\linewidth]{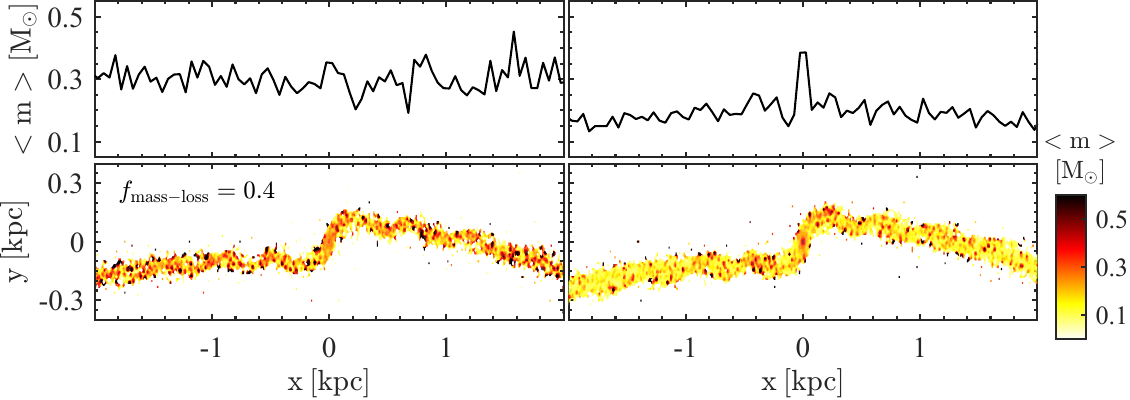}
    \caption{Analysis of the stellar number and mean mass distribution for models at $\RG = 12 \kpc$ with the clusters having lost $40\% $ of their initial mass ($\floss = 0.4$). The left column shows the results for the S0 model, and the right column displays the S1 model. The top two rows show the spatial distribution of the number of luminous stars along the tails (first row) and the corresponding spatial number distributions (second row). The pronounced maximum at $x=0$ ($x$ being the coordinate along the tidal tail) is the cluster and the maximum near $x\approx \pm 0.5,~ \pm 1.0,~ \pm 1.5$ is the K{\"u}pper epicyclic overdensities \citep{Kupper2008, Kupper2010}. The color bar indicates $\log_{10} N$. The bottom two rows show the distribution of the mean stellar mass and the corresponding spatial distribution, with the color bar indicating the mean stellar mass, $<m>$. }
    \label{fig:RG12-f40}
\end{figure*}

\begin{figure*}
    \includegraphics[width=0.87\linewidth]{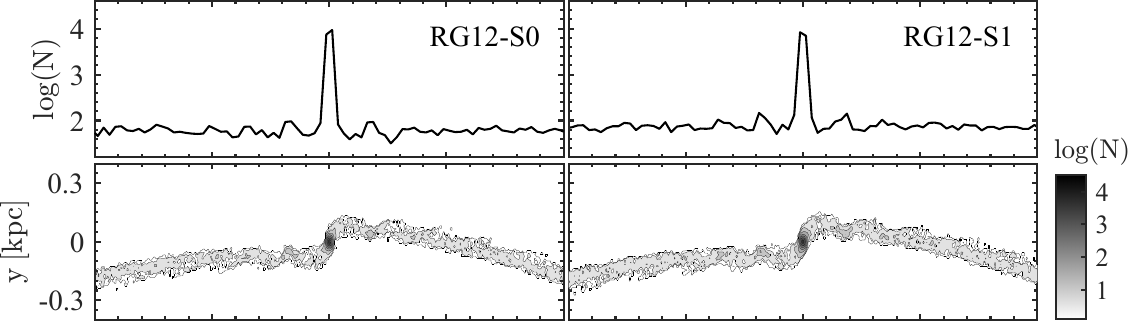}
    \includegraphics[width=0.87\linewidth]{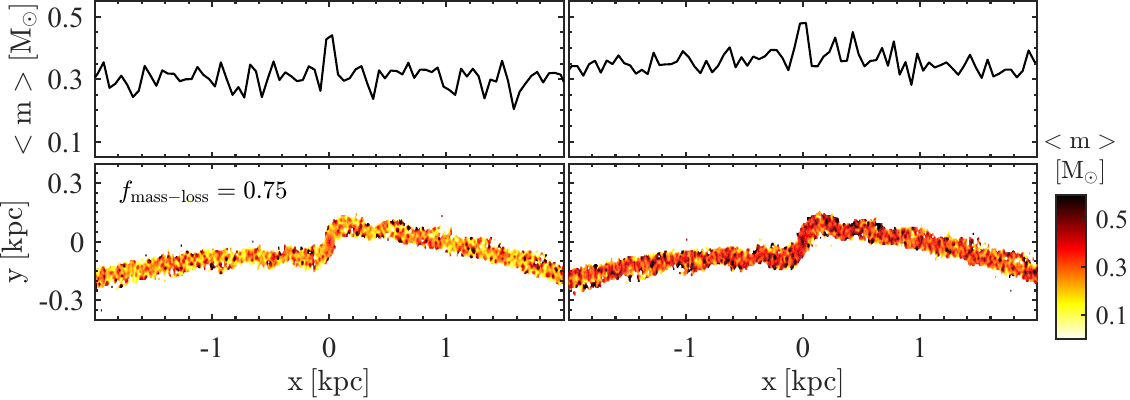}
    \caption{Same as \figref{fig:RG12-f40}, but at $\floss = 0.75$.}
    \label{fig:RG12-f75}
\end{figure*}

\begin{figure}
    \centering
    \includegraphics[width=1.0\linewidth]{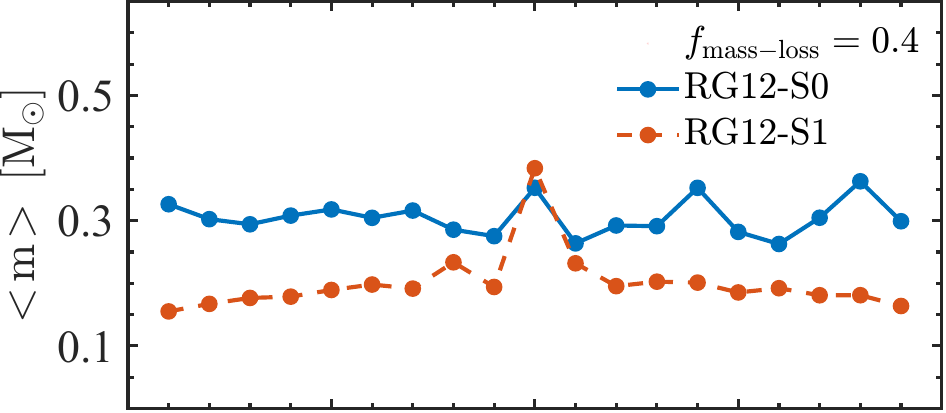}
    \includegraphics[width=1.0\linewidth]{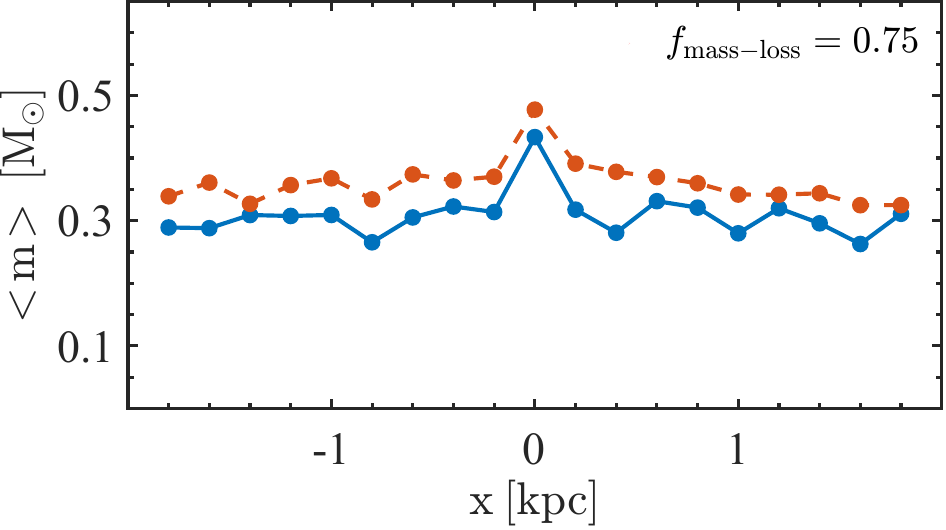}
    \caption{Radial variation of $\mm$ of luminous stars along cluster tails for both S0 and S1 models at $\RG = 12 \kpc$. Distributions are shown at $\floss = 0.4$ and $0.75$ in the top and bottom panels, respectively. The $\mm$ is calculated in 200 pc radial bins. The blue dotted lines show the S0 model, and the orange dotted-dashes show the S1 model.}
    \label{fig:MeanMass-RG12}
\end{figure}

\begin{figure}
    \centering
    \includegraphics[width=1.0\linewidth]{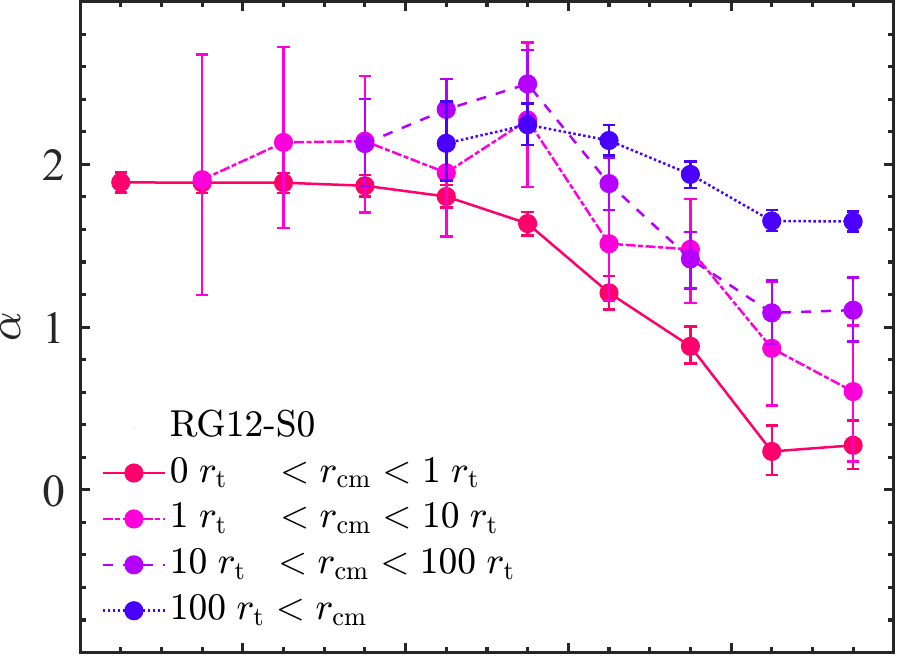}
    \includegraphics[width=1.0\linewidth]{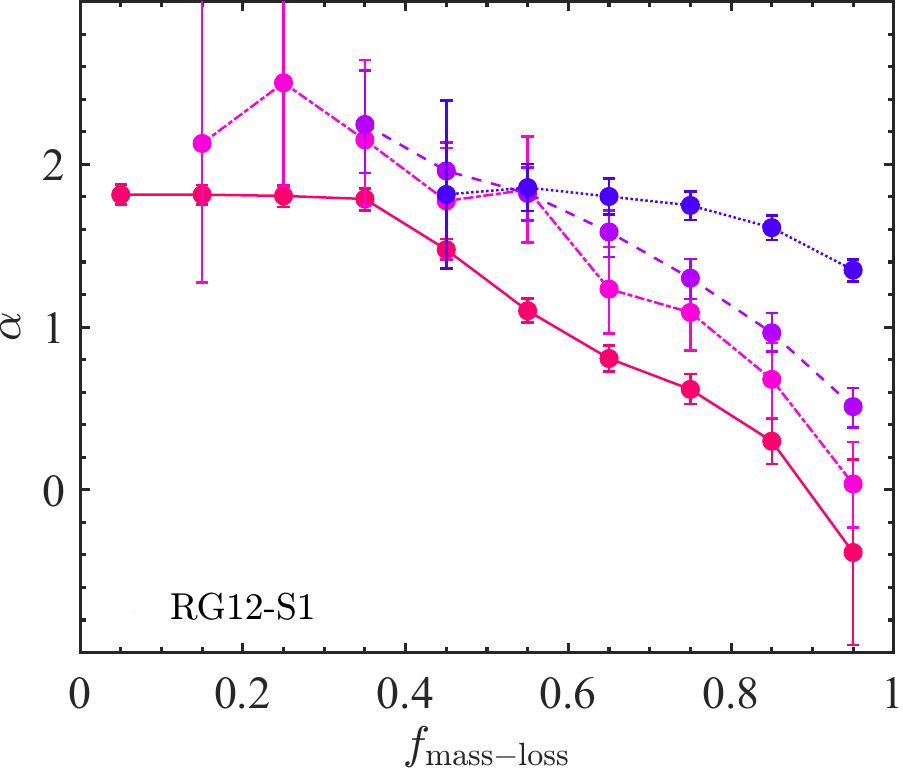}
    \caption{Evolution of the mass function slope as a function of $\floss$ for models at $\RG = 12 \kpc$. $\alpha$ is calculated for luminous stars with masses 0.3 to 0.8 $\Msun$ (eq. \ref{IMF-canonic}). A colour gradient from red to blue represents radial regions from cluster core to tail edges. This visualisation demonstrates the spatial and temporal variation of $\alpha$ across the cluster's evolution.}
    \label{fig:MassFunction-RG12}
\end{figure}

\subsection{Impact of PMS on Tidal Tail Structure}

As illustrated in \figref{fig:initial-MeanMass-contour}, star clusters with PMS initially position lower-mass stars predominantly in their outer regions. Therefore, one can expect that they lose low-mass stars early on, forming tidal tails with a bottom-heavy stellar mass function. Meanwhile, massive stars remain bound longer, later evolving into remnants that form a dense core and accelerate mass loss in later stages \citep{Rostami2024}. Thus, PMS impacts both the early and long-term dynamical evolution and the tail composition of star clusters.

\subsubsection{Overdensities and tail structure modifications}

\figref{fig:RG12-f40} and \figref{fig:RG12-f75} illustrate the stellar number density and mean mass along tidal tails of clusters at $\RG = 12 \kpc$ with low and high PMS. The contrast between S0 and S1 models is clearest during early evolution. In both cases, overdensities form as a natural outcome of epicyclic motions \citep{Kupper2008}, but PMS modifies their strength and appearance. S1 clusters produce denser overdensities with smoother transitions in surrounding regions, whereas S0 clusters show sharper gradients. As clusters evolve and lose mass, these differences weaken, pointing to a gradual homogenization of tail structures over time.

At $\floss = 0.4$, S0 clusters develop symmetric tails that reach a total length of $\approx 4 \kpc$, while S1 clusters produce shorter tails of $\approx 3 \kpc$. This disparity arises largely from the different timescales each cluster type requires to reach the same $\floss$. Despite their shorter extent, S1 tails are noticeably broader.
To quantify this effect, we divide the 2D projected tails into consecutive $50$ pc segments along their length and measure, for each segment, the full width where the stellar surface density exceeds 5 $\mathrm{pc}^{-2}$. The mean tail width is then obtained by averaging over all segments. At $R_{\rm G} = 12$ kpc and $f_{\mathrm{mass-loss}} = 0.4$, the S1 tails are approximately 17\% wider than the S0 tails. This difference decreases to about 10\% at the later evolutionary stage of $f_{\mathrm{mass-loss}} = 0.75$.
This broadening stems from the early escape of low-mass stars, which are ejected with relatively high velocities and disperse more widely around the orbit. As clusters continue to evolve, the tail widths of S0 and S1 models converge within a few $\kpc$ around the clusters, indicating that PMS-driven differences in tail morphology are strongest at early stages but fade over time.

The temporal evolution of $\mm$ along the tails (\figref{fig:MeanMass-RG12}) highlights contrasting behavior between S0 and S1 clusters. In central regions, both models show a gradual increase in $\mm$, but the tails diverge: S0 clusters display a slight decline, while S1 clusters show a marked increase. This reflects the sequential escape of stars in S1, starting with low-mass stars and progressing to higher masses. By $\floss = 0.75$, the differences in $\mm$ between cluster centers and tails are reduced, suggesting a diminishing impact of PMS on tail composition. The contour maps in \figref{fig:RG12-f40} and \figref{fig:RG12-f75} provide a complementary view, showing how PMS reshapes the spatial distribution of stellar masses as clusters evolve.

\subsubsection{Mass function slope ($\alpha$) trends along tails}

\figref{fig:MassFunction-RG12} shows the evolution of the MF slope ($\alpha$) across $\floss$ milestones and radial bins (scaled by $\rt$) for clusters at $\RG = 12 \kpc$. We define the stellar mass function as

\begin{equation}
\frac{\mathrm{d}N(m)}{\mathrm{d}m} \propto m^{-\alpha} 
\label{IMF-canonic}
\end{equation}
where $\alpha$ is the mass–function slope measured for luminous stars in the mass range 0.3–0.8 $\Msun$. 

Both S0 and S1 models show a general decrease in $\alpha$ from the cluster core to the tails, though outer regions tend to stabilize, with mid-tail regions sometimes steepening until $\floss = 0.5$, while tail edges remain near-primordial slope \citep{Webb2022}. S1 clusters show steeper changes in the inner regions but flatter $\alpha$ in the outskirts. Early tail formation in S1 models yields higher $\alpha$ (more bottom-heavy MFs), whereas a broader stellar mass range reveals a rapid early rise in $\alpha$ due to preferential low-mass star loss.

At early evolutionary stages, the MF slope ($\alpha$) in the tidal streams can temporarily exceed the primordial value because the first escapers are preferentially low-mass stars originating from the cluster outskirts, leading to a steeper stream MF. As the cluster evolves dynamically, mass segregation drives massive stars toward the cluster center, and subsequent escape processes progressively populate the streams with a broader stellar mass range, causing the MF slope to decrease over time. Eventually, all radial bins evolve below the primordial slope due to the combined effects of stellar evolution, preferential escape of massive stars, and long-term dynamical mass loss, which continuously modifies the mass distribution of stars populating the streams.

\subsubsection{Evolution and convergence of $<m>$ gradients in tidal tails} 

\begin{figure*}
    \centering
    \includegraphics[scale=1.05]{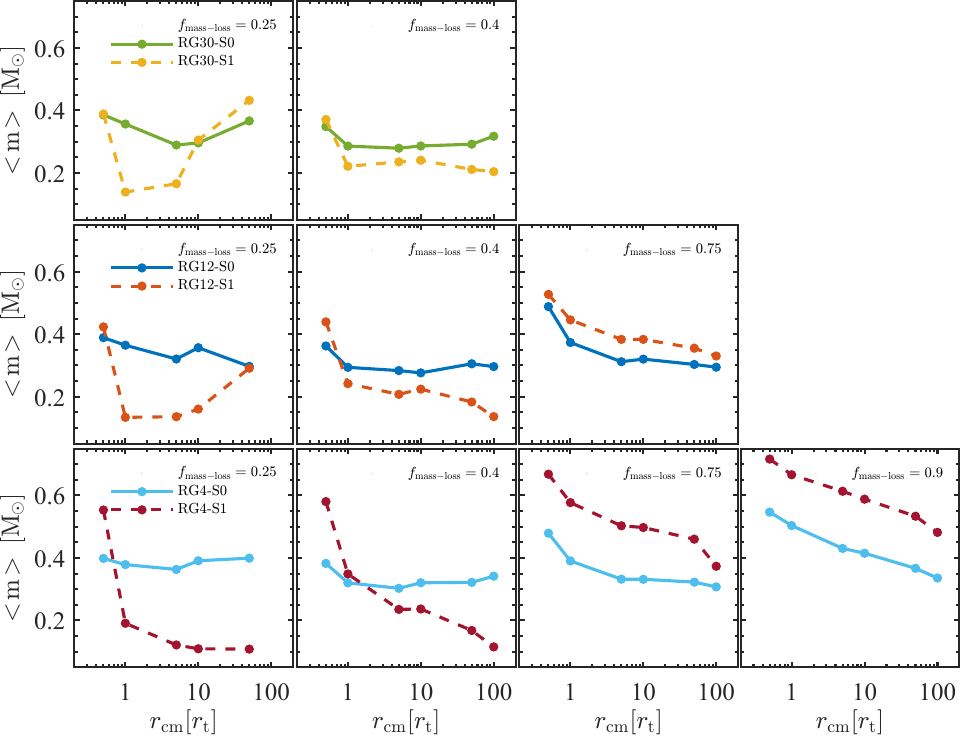}
    \caption{Comparative analysis of mean stellar mass ($<m>$) profiles in tidal tails of low-NKV cluster models. The horizontal axis shows distance from the cluster centre in units of tidal radius ($\rt$), plotted on a semi-logarithmic scale from 0 to 100 $\rt$. The vertical axis shows the mean stellar mass, $<m>$. Columns correspond to different evolutionary stages, expressed by fractional mass loss ($\floss$, see \tabref{table:all}), increasing from left to right. Rows group models at the same Galactocentric radius ($\RG$). Together, the panels trace how stellar masses are distributed along the tails at different stages of cluster evolution.}
    \label{fig:all-lowkick}
\end{figure*}

An extensive investigation of tidal tail structures is shown in \figref{fig:all-lowkick}, which presents the $<m>$ distribution of luminous stars for all low-NKV models using semi-logarithmic radial binning from the cluster centre ($\rcm$). The bottom row highlights temporal evolution at $\RG = 4 \kpc$, where both low- and high-segregated clusters show that S1 tail gradients gradually converge toward S0 values. Yet, S1 models retain slightly higher $<m>$ at late stages. This is due to the late outflow of heavier components in highly segregated clusters. The convergence of the $<m>$ slopes indicates that long-term processes like two-body relaxation and tidal stripping drive uniformity and diminish initial PMS-driven differences.

To track $<m>$ distributions along the tails, we fit Gaussians characterized by dispersion ($\sigma_{<m>}$) and amplitude ($Amp$). \tabref{table:results-gaussian} lists values at $\floss = 0.4, 0.75, 0.9$. $Amp$ generally increases with evolution, more strongly in S0 models, while $\sigma_{<m>}$ diverges: in S0 clusters it decreases over time, but in S1 clusters it grows as the $<m>$ gradient softens. It should be noted that the Gaussian fits were applied to the spatial variation of the mean stellar mass, ($<m>(x)$), measured along the tidal-tail coordinate ($x$), rather than to the masses of individual stars. Specifically, for each snapshot, we compute ($<m>$) within consecutive bins along the tidal tail and analyze its variation as a function of position.

The long-term increase in the $Amp$ parameter in all models reflects the mass segregation process occurring in star clusters, but we expect less changes in S1 models due to the initially mass-segregated formation of stars. On the other hand, for the $\sigma_{<m>}$ scenarios, in S1 clusters we have a high concentration of heavy components at the centre, which peaks $<m>$ value at this region, resulting in a small dispersion. But as clusters evolve and the dispersion of mass moves toward a unified distribution, the $<m>$ flattens, leading to a wider and larger   $\sigma_{<m>}$ . Whereas for S0 models, the initial distribution of mass along the tails is uniform, leading to an extremely wide $\sigma_{<m>}$. As S0 clusters evolve, the dynamical mass segregation leads to more concentration of higher-mass stars at the centre, making a sharper peak of $\mm$ is this region, which gradually shrinks the $\sigma_{<m>}$ value through time.

\begin{table}
    \caption{The values for the dispersion ($\sigma_{<m>}$) and amplitude (Amp) of the fitted Gaussian functions to the distribution of $<m>$ are shown in this table.}
	\centering
	\begin{tabular}{ccccc}
		
  		\hline 	
            \hline
            Model & &     & $\floss$ &     \\
                  & & 0.4 & 0.75     & 0.9 \\
            & & $\sigma_{\mm}$ ($Amp$) & $\sigma_{\mm}$ ($Amp$) & $\sigma_{\mm}$ ($Amp$) \\
            \hline
            
            RG4-S0    & & 191.34 (0.01) & 5.41 (0.05)  & 4.65 (0.07)  \\
            RG4-S1    & & 1.47   (0.16) & 2.78 (0.20)  & 2.95 (0.22)  \\
            \hline
            RG4-S0-K  & & 121.87 (0.00) & 12.82 (0.01) & 4.31 (0.08)  \\
            RG4-S1-K  & & 5.6    (0.13) & 2.94  (0.18) & 3.65 (0.15)   \\
            \hline
            RG12-S0   & & 365.3  (0.00) & 8.00 (0.02)  & 4.12 (0.10)  \\
            RG12-S1   & & 3.3    (0.08) & 5.30 (0.06)  & 5.41 (0.07)  \\
            \hline
            RG30-S0   & & 117.77 (0.03) & -            & -            \\ 
            RG30-S1   & & 5.6    (0.03) & -            & -            \\
                  
            \hline
            \hline
        
	\end{tabular}
	\label{table:results-gaussian}
\end{table}

\begin{figure}
    \centering
    \includegraphics[width=1.0\linewidth]{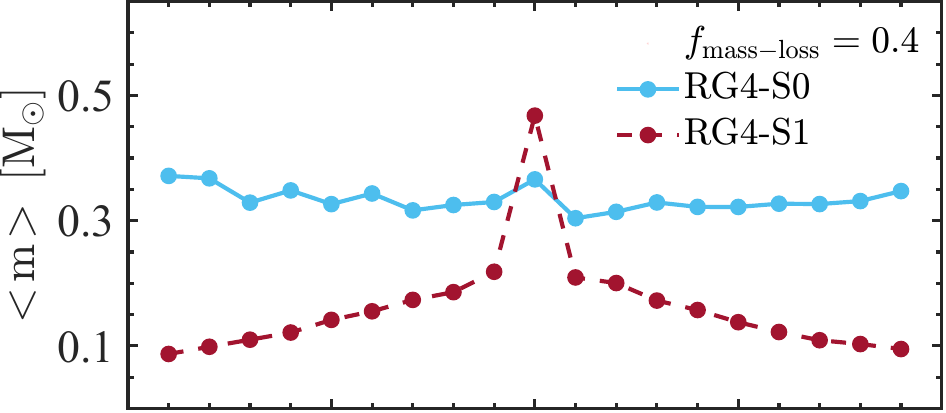}
    \includegraphics[width=1.0\linewidth]{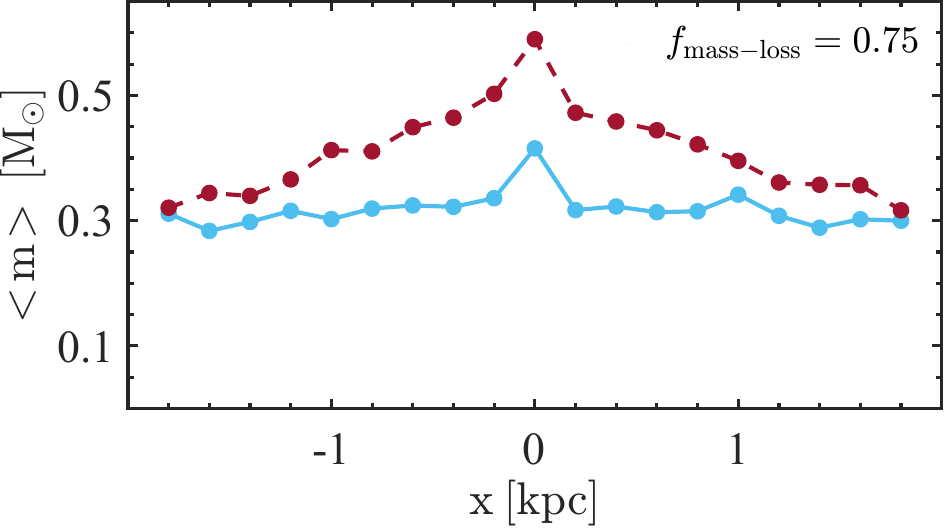}
    \caption{Same as \figref{fig:MeanMass-RG12}, but for models at $\RG = 4 \kpc$. The top and bottom panels show the snapshots of $\floss = 0.4$ and $0.75$, respectively. The cyan dotted lines show the RG4-S0 model, and red dot-dashes show the RG4-S1 model.}
    \label{fig:MeanMass-RG4}
\end{figure}

%--------------------------------------------------------------
\subsection{Impact of $\RG$ on Tidal Tail Structure and PMS Signatures}

The Galactocentric radius strongly shapes cluster evolution and tidal tail properties. In our models, clusters were placed on circular orbits to isolate $\RG$ effects. Clusters at larger $\RG$ lose mass more slowly and evolve over longer timescales, producing wide, diffuse tails. By contrast, those near the Galactic centre undergo faster mass loss and stronger tidal forcing, yielding compact, coherent tails. Thus, the balance between $\RG$, mass loss, and tidal forces creates a spectrum of tail morphologies, from sparse extended structures in the outer Galaxy to dense tails in the inner regions.

When combined with PMS, the impact of $\RG$ becomes clearer. At small radii (e.g., $\RG = 4 \kpc$), PMS effects on $<m>$ are much stronger than at $\RG = 12 \kpc$, both early and late in the evolution. Increasing $\RG$ therefore diminishes PMS signatures in tidal tails. The Gaussian fits of $<m>$ at $\floss = 0.75$ (\tabref{table:results-gaussian}) confirm this trend: clusters at smaller $\RG$ show higher $Amp$ and lower $\sigma_{<m>}$.

Using all low-NKV models, we further examined $<m>$ gradients along leading and trailing tails, with semi-logarithmic bins scaled to each cluster’s $\rt$ (\figref{fig:all-lowkick}). The bin limits are set at 0, 0.5, 1, 5, 10, 50, and 100 $\rt$. The results for each radial bin are plotted at the bin's upper boundary. This analysis illustrates how PMS and $\RG$ jointly shape long-term dynamical evolution, showing how internal dynamics and external tidal forces interact to set tail structure across different galactic environments.

Diving deeper into the \figref{fig:all-lowkick}, at early $\floss = 0.25$, inner–outer $<m>$ differences are stronger at small $\RG$, reflecting the early mass segregation profile, while clusters at larger $\RG$ show weaker contrasts due to longer evolutionary timescales. Outer tails at large radii are more chaotic and sparse, while the $<m>$ gradients along $\rcm$ smooth out over time. At $\floss = 0.25$–0.4, S0 and S1 profiles intersect with different slopes, but at later stages the slopes converge, though average $<m>$ offsets remain. This alignment occurs earlier for clusters at larger $\RG$. Over time, S1 profiles flatten, whereas S0 trends steepen.

\begin{figure*}
    \centering
    \includegraphics[width=0.44\linewidth]{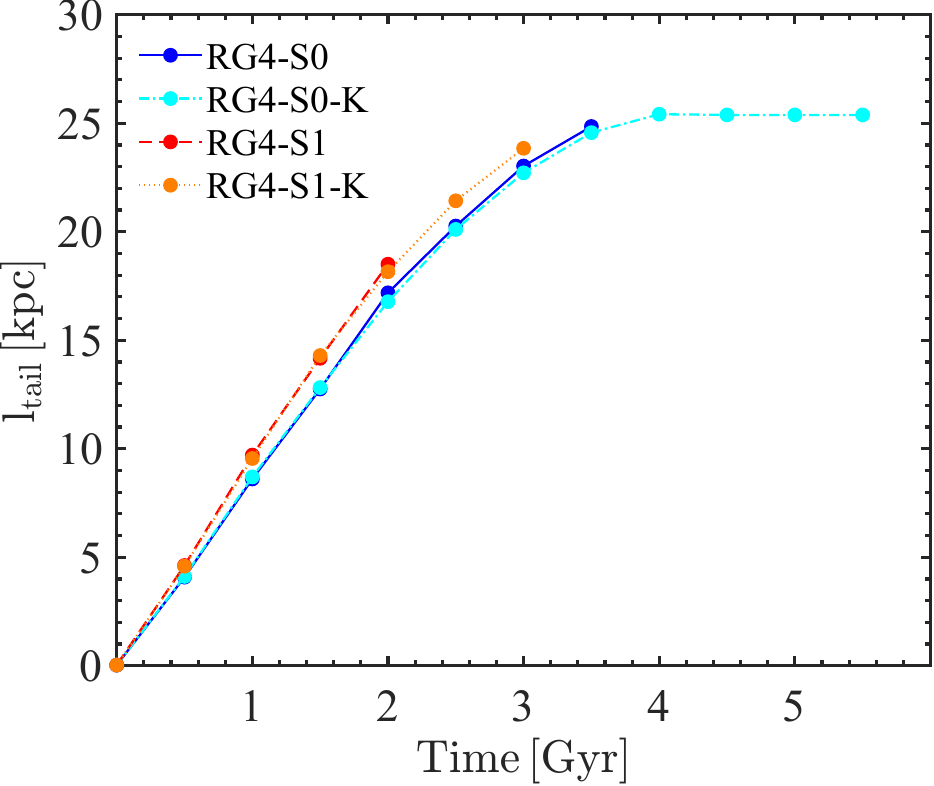}
    \includegraphics[width=0.387\linewidth]{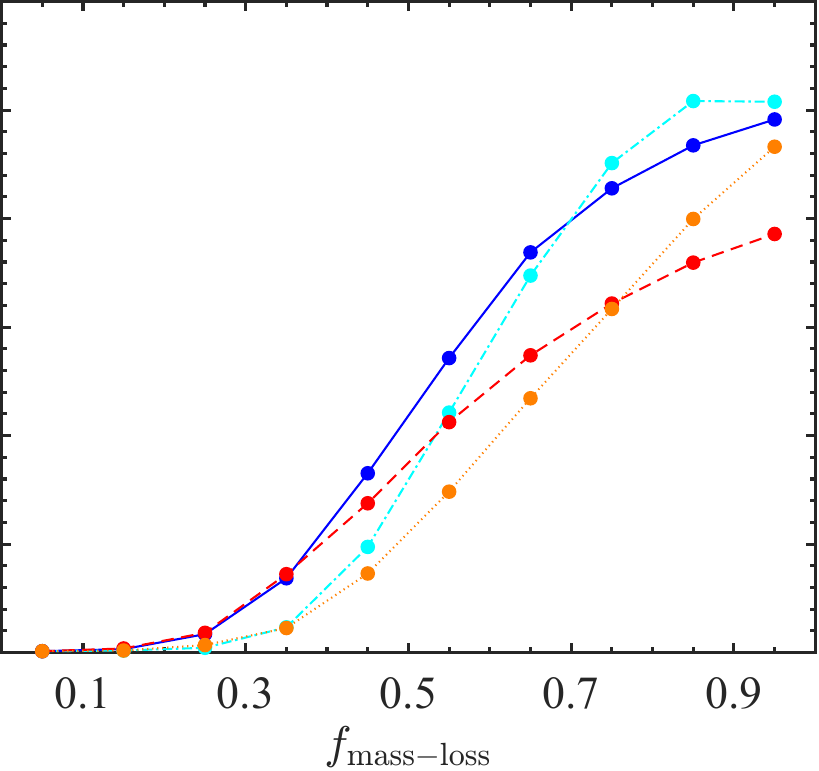}
    \caption{The total length of the leading and trailing tails of star clusters evolving with time (left panel) and $\floss$ (right panel) for all 4 models placed at $\RG = 4 \kpc$ until their dissolution times.}
    \label{fig:tail-length}
\end{figure*}

\begin{figure}
    \centering
    \includegraphics[width=1.0\linewidth]{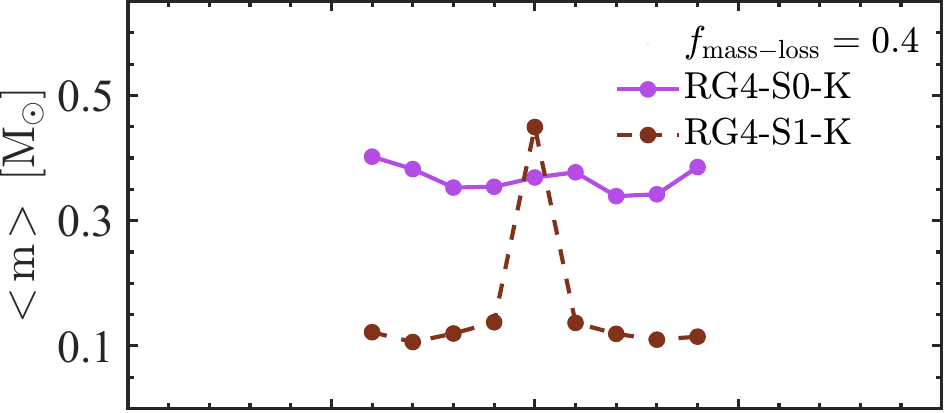}
    \includegraphics[width=1.0\linewidth]{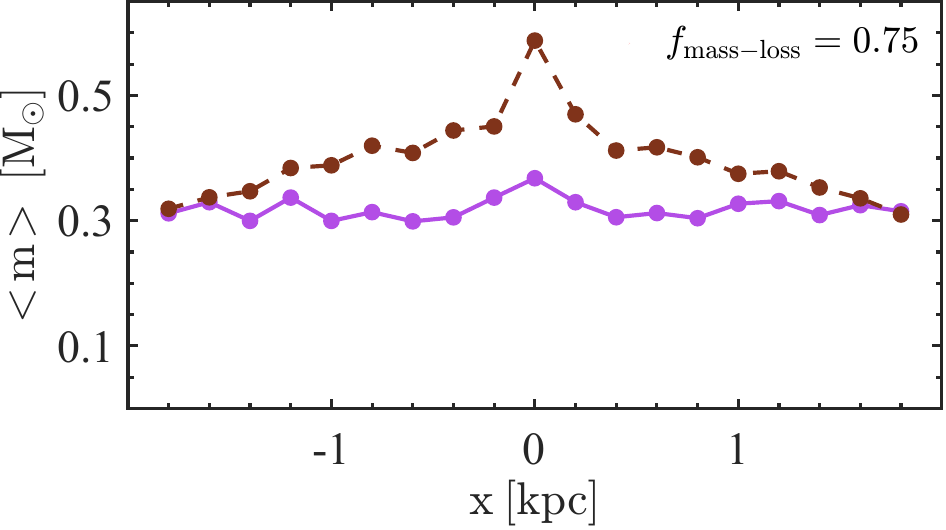}
    \caption{Same as \figref{fig:MeanMass-RG4}, but for models with high NKV at $\RG = 4 \kpc$. The top and bottom panels show the $\floss = 0.4$ and $0.75$, respectively. The magenta dotted lines show the RG4-S0-K model, and brown dot-dashes show the RG4-S1-K model.}
    \label{fig:MeanMass-RG4-K}
\end{figure}

%--------------------------------------------------------------

\subsection{Influence of NKV on Tidal Tail Evolution}

The NKV of stellar remnants strongly affects star cluster evolution by controlling the retention of BHs. A bound BH population forms a central subsystem that heats the cluster, enhancing stellar escape and accelerating dynamical evolution \citep{Rostami2024}. \figref{fig:tail-length} illustrates the combined effects of PMS and NKV on the cumulative tidal tail length $\ltail$ for clusters at $\RG = 4 \kpc$, measured as the orbital span between the most distant leading and trailing segments containing at least 20 luminous stars within 10 pc.

The temporal evolution of $\ltail$ (left panel) shows that S0 clusters follow similar trajectories under both low and high NKV, as do S1 clusters. However, S1 models display faster tail growth, driven by the early escape of low-mass stars from their segregated outskirts. The $\floss$-dependent evolution (right panel) reveals that low-NKV clusters form longer tails at early times, since supernova-induced mass loss inflates $\floss$ without strongly altering stellar velocities. As clusters lose mass and approach $\floss \approx 0.75$, evaporation rates stabilize and $\ltail$ converges across NKV values, though S0 clusters consistently retain  about 1.4 times longer tails than S1 model, underscoring the lasting role of PMS.

The impact of NKV on $<m>$ distributions is shown in \figref{fig:MeanMass-RG4-K}. Compared with the low-NKV case (\figref{fig:MeanMass-RG4}), high NKV does not significantly change $<m>$ differences between S0 and S1 tails. Its main influence appears at early stages ($\floss < 0.4$), when reduced tail length produces flatter $<m>$ gradients. At later times, as $\ltail$ extends in both scenarios, $<m>$ profiles converge and NKV effects fade. Overall, while NKV shapes early tail growth and structure, its influence weakens over time, whereas PMS continues to imprint long-term differences in tidal tails.

%%%%%%%%%%%%%%%%%%%%%%%%%%%%%%%%%%%%%%%%%%%%%%%%%%%%%%%%%%%%%%%%%%%%%
%%%%%%%%%%%%%%%%%%%%%%%%%%%%%%%%%%%%%%%%%%%%%%%%%%%%%%%%%%%%%%%%%%%%%

\section{Conclusions and Discussion}

In this study, we investigated how PMS, Galactocentric distance, and stellar remnant retention influence the dynamical evolution of star clusters and their tidal tails. Using direct N-body simulations, we identified the following key results:
\begin{itemize}
 \item

PMS accelerates early cluster expansion by concentrating massive stars in the cluster core, such that their early mass loss efficiently reduces the depth of the central potential. The resulting expansion leads to the preferential loss of low-mass stars. This rapid evaporation leads to the formation of tidal tails with bottom-heavy stellar mass functions, especially at smaller Galactocentric radii ($R_{\rm G}$). Primordially segregated clusters develop longer, wider, and denser tidal tails than their non-segregated counterparts.

\item
At inner radii (e.g. $R_{\rm G}=4$ kpc), the influence of PMS is strongest: tails are more populated and their morphology shows clear signatures of PMS. At larger $R_{\rm G}$, clusters lose mass more slowly and form more extended but diffuse tidal tails.

\item  
PMS fingerprints are most visible at early and intermediate evolutionary stages.  At fixed mass-loss fractions, initially mass-segregated models show tails up to 1.4 times longer than non-segregated models.
Differences in mean stellar mass, $<m>$, along the tails between segregated and non-segregated clusters diminish over time, eventually converging. This indicates that PMS effects on tidal structures gradually fade as clusters evolve toward dissolution.

\item 
Black hole and stellar remnant retention, shaped by natal kick velocities, exerts only a minor influence on tail morphology and mean stellar mass distribution. The primary drivers of tidal tail evolution remain PMS and Galactocentric distance.

\item 
PMS strongly shapes the early dynamical history of clusters and the morphology of their tidal tails. Over time, however, these initial conditions leave progressively weaker imprints, highlighting that time erases the dynamical memory of PMS.  Therefore, while tidal tails at young ages may retain clear PMS fingerprints, at late stages the signatures become blurred, complicating their observational identification.

\end{itemize}

%%%%%%%%%%%%%%%%%%%%%%%%%%%%%%%%%%%%%%%%%%%%%%%%%%%%%%%%%%%%%%%%%%%%%
%%%%%%%%%%%%%%%%%%%%%%%%%%%%%%%%%%%%%%%%%%%%%%%%%%%%%%%%%%%%%%%%%%%%%

% Bibliography
\section*{Acknowledgements}

HH is grateful to the staff at the Helmholtz-Institut f{\"u}r Strahlen- und Kernphysik (HISKP) and Argelander Institut f{\"u}r Astronomie (AIfA) for their hospitality and acknowledges financial support from the Stellar Populations and Dynamics (SPODYR) group at the University of Bonn. HH also acknowledges support from the Center for International Scientific Studies \& Collaboration (CISSC),  Ministry of Science, Research and Technology of Iran. AHZ acknowledges support from the Alexander von Humboldt Foundation. PK thanks the Deutscher Akademischer Austauschdienst (DAAD) Bonn-Prague Eastern European Exchange Program.

\section*{Data availability}
The data underlying this paper are available in the paper.

\bibliographystyle{mnras}
\bibliography{references}

%%%%%%%%%%%%%%%%%%%%%%%%%%%%%%%%%%%%%%%%%%%%%%%%%%%%%%%%%%%%%%%%%%%%%
%%%%%%%%%%%%%%%%%%%%%%%%%%%%%%%%%%%%%%%%%%%%%%%%%%%%%%%%%%%%%%%%%%%%%

\appendix

\bsp
\label{lastpage}
\end{document}